\documentclass[twocolumn,showpacs,prc,amsmath,amssymb,superscriptaddress]{revtex4}

\usepackage{graphicx}
\usepackage{dcolumn}
\usepackage{bm}
\usepackage{ulem}
\usepackage{epsfig}
\usepackage{fancyhdr}
\usepackage{amsmath}
\usepackage{amsfonts}
\usepackage{natbib}
\usepackage{amsmath}
\begin{document}

\preprint{}

\title{N and Z odd-even staggering in Kr + Sn collisions at Fermi energies}

\author{S.~Piantelli}
\affiliation{Sezione INFN di Firenze, Via G. Sansone 1, 
     I-50019 Sesto Fiorentino, Italy}

\author{G.~Casini}
\affiliation{Sezione INFN di Firenze, Via G. Sansone 1, 
     I-50019 Sesto Fiorentino, Italy}

\author{P.R.~Maurenzig}
\affiliation{Sezione INFN di Firenze, Via G. Sansone 1, 
     I-50019 Sesto Fiorentino, Italy}
\affiliation{Dipartimento di Fisica, Univ. di Firenze, 
     Via G. Sansone 1, I-50019 Sesto Fiorentino, Italy}

\author{A.~Olmi}
\thanks{corresponding author} \email[e-mail:]{olmi@fi.infn.it}
\affiliation{Sezione INFN di Firenze, Via G. Sansone 1, 
     I-50019 Sesto Fiorentino, Italy}

\author{S.~Barlini}
\affiliation{Sezione INFN di Firenze, Via G. Sansone 1, 
     I-50019 Sesto Fiorentino, Italy}
\affiliation{Dipartimento di Fisica, Univ. di Firenze, 
     Via G. Sansone 1, I-50019 Sesto Fiorentino, Italy}

\author{M.~Bini}
\affiliation{Sezione INFN di Firenze, Via G. Sansone 1, 
     I-50019 Sesto Fiorentino, Italy}
\affiliation{Dipartimento di Fisica, Univ. di Firenze, 
     Via G. Sansone 1, I-50019 Sesto Fiorentino, Italy}

\author{S.~Carboni}
\affiliation{Sezione INFN di Firenze, Via G. Sansone 1, 
     I-50019 Sesto Fiorentino, Italy}
\affiliation{Dipartimento di Fisica, Univ. di Firenze, 
     Via G. Sansone 1, I-50019 Sesto Fiorentino, Italy}

\author{G.~Pasquali}
\affiliation{Sezione INFN di Firenze, Via G. Sansone 1, 
     I-50019 Sesto Fiorentino, Italy}
\affiliation{Dipartimento di Fisica, Univ. di Firenze, 
     Via G. Sansone 1, I-50019 Sesto Fiorentino, Italy}

\author{G.~Poggi}
\affiliation{Sezione INFN di Firenze, Via G. Sansone 1, 
     I-50019 Sesto Fiorentino, Italy}
\affiliation{Dipartimento di Fisica, Univ. di Firenze, 
     Via G. Sansone 1, I-50019 Sesto Fiorentino, Italy}

\author{A.A.~Stefanini}
\affiliation{Sezione INFN di Firenze, Via G. Sansone 1, 
     I-50019 Sesto Fiorentino, Italy}
\affiliation{Dipartimento di Fisica, Univ. di Firenze, 
     Via G. Sansone 1, I-50019 Sesto Fiorentino, Italy}

\author{S.~Valdr\`e}
\affiliation{Sezione INFN di Firenze, Via G. Sansone 1, 
     I-50019 Sesto Fiorentino, Italy}
\affiliation{Dipartimento di Fisica, Univ. di Firenze, 
     Via G. Sansone 1, I-50019 Sesto Fiorentino, Italy}

\author{R.~Bougault}
\affiliation{LPC, IN2P3-CNRS, ENSICAEN et Universit\'{e} de Caen, 
 F-14050 Caen-Cedex, France}

\author{E.~Bonnet}
\affiliation{GANIL, CEA/DSM-CNRS/IN2P3, B.P. 5027, F-14076 Caen cedex, France}

\author{B.~Borderie}
\affiliation{Institut de Physique Nucl\'{e}aire, CNRS/IN2P3, 
      Universit\'{e} Paris-Sud 11, F-91406 Orsay cedex, France}

\author{A.~Chbihi}
\affiliation{GANIL, CEA/DSM-CNRS/IN2P3, B.P. 5027, F-14076 Caen cedex, France}

\author{J.D.~Frankland}
\affiliation{GANIL, CEA/DSM-CNRS/IN2P3, B.P. 5027, F-14076 Caen cedex, France}

\author{D.~Gruyer}
\affiliation{GANIL, CEA/DSM-CNRS/IN2P3, B.P. 5027, F-14076 Caen cedex, France}

\author{O.~Lopez}
\affiliation{LPC, IN2P3-CNRS, ENSICAEN et Universit\'{e} de Caen, 
 F-14050 Caen-Cedex, France}

\author{N.~Le~Neindre}
\affiliation{LPC, IN2P3-CNRS, ENSICAEN et Universit\'{e} de Caen, 
 F-14050 Caen-Cedex, France}

\author{M.~P\^{a}rlog}
\affiliation{LPC, IN2P3-CNRS, ENSICAEN et Universit\'{e} de Caen, 
 F-14050 Caen-Cedex, France}
\affiliation{Horia Hulubei, National Institute of Physics and 
    Nuclear Engineering, RO-077125 Bucharest-M\u{a}gurele, Romania}

\author{M.F.~Rivet}
\affiliation{Institut de Physique Nucl\'{e}aire, CNRS/IN2P3, 
      Universit\'{e} Paris-Sud 11, F-91406 Orsay cedex, France}

\author{E.~Vient}
\affiliation{LPC, IN2P3-CNRS, ENSICAEN et Universit\'{e} de Caen, 
 F-14050 Caen-Cedex, France}

\author{E.~Rosato}
\affiliation{Sezione INFN di Napoli and Dip. di Fisica,
    Univ. di Napoli ``Federico II'', I 80126 Napoli, Italy}

\author{G.~Spadaccini}
\affiliation{Sezione INFN di Napoli and Dip. di Fisica,
    Univ. di Napoli ``Federico II'', I 80126 Napoli, Italy}

\author{M.~Vigilante}
\affiliation{Sezione INFN di Napoli and Dip. di Fisica,
    Univ. di Napoli ``Federico II'', I 80126 Napoli, Italy}

\author{M.~Bruno}
\affiliation{Sezione INFN di Bologna and Dip. di Fisica, Univ. di Bologna,
             40126 Bologna, Italy}

\author{T.~Marchi}
\affiliation{INFN Laboratori Nazionali di Legnaro, 
      Viale dell'Universit\'{a} 2, 35020 Legnaro (Padova) Italy}

\author{L.~Morelli}
\affiliation{Sezione INFN di Bologna and Dip. di Fisica, Univ. di Bologna,
             40126 Bologna, Italy}

\author{M.~Cinausero}
\affiliation{INFN Laboratori Nazionali di Legnaro, 
      Viale dell'Universit\'{a} 2, 35020 Legnaro (Padova) Italy}

\author{M.~Degerlier}
\affiliation{Nevsehir University Science and Art Faculty, Physics
Department, Nevsehir, Turkey}

\author{F.~Gramegna}
\affiliation{INFN Laboratori Nazionali di Legnaro, 
      Viale dell'Universit\'{a} 2, 35020 Legnaro (Padova) Italy}

\author{T.~Kozik}
\affiliation{Jagiellonian University, Institute of Nuclear Physics 
     IFJ-PAN, PL-31342 Krak\'{o}w, Poland}

\author{T.~Twar\'{o}g}
\affiliation{Jagiellonian University, Institute of Nuclear Physics 
     IFJ-PAN, PL-31342 Krak\'{o}w, Poland}

\author{R.~Alba}
\affiliation{INFN Laboratori Nazionali del Sud, 
             Via S.Sofia 62, 95125 Catania, Italy}

\author{C.~Maiolino}
\affiliation{INFN Laboratori Nazionali del Sud, 
             Via S.Sofia 62, 95125 Catania, Italy}

\author{D.~Santonocito}
\affiliation{INFN Laboratori Nazionali del Sud, 
             Via S.Sofia 62, 95125 Catania, Italy}

\collaboration{FAZIA Collaboration}
\noaffiliation

\date{\today}

\begin{abstract}
The odd-even staggering of the yield of final reaction products has been studied
as a function of proton ($Z$) and neutron $(N)$ numbers for the collisions 
$^{84}$Kr+$^{112}$Sn and $^{84}$Kr+$^{124}$Sn at 35 MeV/nucleon,
in a wide range of elements (up to $Z\approx20$).
The experimental data show that staggering effects rapidly decrease with 
increasing size of the fragments.
Moreover the staggering in $N$ is definitely larger than the one in $Z$.
Similar general features are qualitatively reproduced by the GEMINI code.
Concerning the comparison of the two systems, the staggering in $N$ is 
in general rather similar, being slightly larger only for 
the lightest fragments produced in the $n$-rich system.
In contrast the staggering in $Z$, although smaller than that in $N$, 
is sizably larger for the $n$-poor system with respect to the $n$-rich one.  
\end{abstract}


\maketitle

\section{INTRODUCTION}
\label{sec:Introduction}

The odd-even staggering in the yields of reaction products is a feature 
that has been observed since many years in the charge distributions of a 
large variety of nuclear reactions.
This phenomenon was extensively studied in relation to fission fragments 
of actinide nuclei (see, e.g., \cite{Tracy72,Tsekhanovich99,Schmidt01,Naik07}
and references therein), where it was attributed to pairing effects in the
nascent fragments.

Odd-even staggering was observed also in light fragments produced by
fragmentation or spallation at relativistic energies (see, e.g.,
\cite{Poskanzer71,Zeitlin97,Ricciardi04,Napolitani07}) and more 
recently even in heavy ion collisions at Fermi energies (15$\alt$E/A$\alt$50
MeV/nucleon) \cite{Yang99,Winchester00,Geraci04,Lombardo11,Casini12}.
The study of odd-even effects has gained renewed interest from this last finding.
In fact, in order to study the symmetry energy \cite{colonna05,su11,raduta07}, one
needs to reliably estimate the primary isotopic distributions of fragments
and this is possible only if the effects of secondary decays are small or
sufficiently well understood. 

Usually the staggering consists in even-$Z$ fragments presenting systematically
higher yields with respect to the neighboring odd-$Z$ ones. 
When isotopic identification is achieved (as in spectrometer-based experiments),
additional features emerge: for example, fragments with $N\!=\!Z$ show a
particularly strong staggering, while fragments with odd difference
$N-Z$ present a reverse staggering (``anti-staggering''), favoring the production 
of fragments with odd $Z$ \cite{Ricciardi04,DAgostino11,winkenbauer_arXiv1303}. 
Moreover, if systems with different $N/Z$ are compared, the $n$-poor system shows
an enhanced staggering in the charge distribution with respect to the $n$-rich one
\cite{Yang99,DAgostino11,Lombardo11},
while the opposite is observed for the N distribution \cite{Lombardo11}.  

In low-energy heavy-ion collisions, the odd-even staggering may be a
signature of nuclear structure effects in the reaction mechanism, if part 
of the reaction proceeds through very low excitation energies \cite{Ademard11}.
In collisions at intermediate (or Fermi) energies the preferred interpretation
is that structure effects are restored in the final products of hot decaying
nuclei and that the odd-even staggering depends 
- in a complex and presently not very well understood way - 
on the structure of the nuclei produced near the end of the 
evaporation chain  \cite{Ricciardi04,DAgostino11,DAgostino12}.
At present, no theoretical model exists that is able to reproduce all the details
of the observed staggering, although some general characteristics are reproduced.
For example, in \cite{DAgostino11} a staggering effect is observed in events
simulated with the {\sc gemini} code \cite{CharityPhysRevC.82.014610},
where staggering originates from the mass parametrization
that includes a pairing contribution \cite{MollerNix1981}, fading out with 
increasing excitation energy and spin.
A comparison with the results of {\sc ismm} \cite{Souza03} is presented 
in \cite{winkenbauer_arXiv1303}, where staggering is attributed to a 
pairing-dependent term, rapidly oscillating as a function of Z, that 
affects an otherwise smooth distribution.

In this work we present an analysis of the data taken by the FAZIA 
Collaboration \cite{FAZIAcoll} in the collisions $^{84}$Kr +$^{112}$Sn 
(henceforth ``$n$-poor'' system) and $^{84}$Kr+$^{124}$Sn 
(``$n$-rich'' system) at a bombarding energy of 35 MeV/nucleon. 
The odd-even staggering effects are investigated as a function of 
atomic number ($Z$ staggering) and neutron number ($N$ staggering) 
for the two colliding systems.
Some comparisons with the results of {\sc gemini} and with other experimental 
data available in literature are presented too. 

\section{SETUP AND EXPERIMENTAL RESULTS}
\label{sec2}

The experiment was performed at the Superconducting Cyclotron of LNS 
(Laboratori Nazionali del Sud) of INFN in Catania. 
A pulsed beam of $^{84}$Kr at 35 MeV/nucleon was used to bombard two targets 
of $^{112}$Sn (areal density 415 $\mu$g/cm$^2$) and
$^{124}$Sn (areal density 600 $\mu$g/cm$^{2}$). 
Reaction products were detected in a Si-Si-CsI(Tl) telescope of the FAZIA
Collaboration  (thicknesses: 300 $\mu$m, 500 $\mu$m and 10 cm, respectively), 
covering the angular range 
between 4.8$^{\circ}$ and 6$^{\circ}$, close to the grazing angles of 
the two reactions (4.1$^{\circ}$ for the $n$-poor and 4.0$^{\circ}$ for the 
$n$-rich system). 
The same set of data was analyzed also in a recent paper 
\cite{Barlini13}, where the good performances of the FAZIA
telescope in terms of charge and mass identification capability were used to
investigate the isospin transport by means of fragments isotopically resolved 
up to $Z=20$.  
More details on the experimental setup can be found 
in \cite{Carboni12,Barlini13}
while the performances of the FAZIA telescopes are illustrated in 
\cite{Bardelli2009501,Bardelli2011272,Carboni12,LeNeindre13,Barlini13NIM}.

\begin{figure}[t]
\begin{center}
\includegraphics[width=9cm] {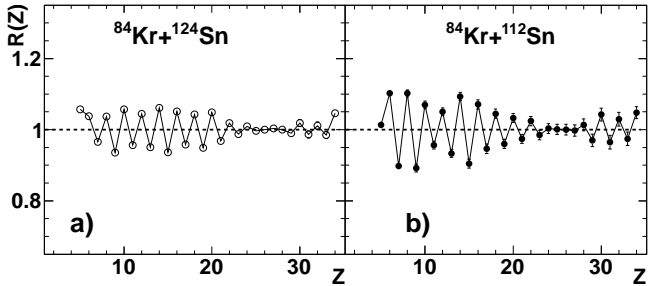} 
\end{center}
\caption{Staggering as a function of $Z$, highlighted by the ratio $R(Z)$ 
  for the system (a) $^{84}$Kr+$^{124}$Sn and (b) $^{84}$Kr+$^{112}$Sn, 
  both at 35 MeV/nucleon. Bars indicate statistical errors.}
\label{fig1}
\end{figure}

The present analysis concerns ions identified with the $\Delta E -E$ technique, 
as it was done in Ref. \cite{Barlini13}.
The data were acquired in singles, so a characterization of the 
centrality of the collisions is not possible. 
However, as explained on the basis of Fig. 2 in \cite{Barlini13}, from the 
accessible phase space region one can expect that most detected products are 
either quasi-projectile residues ($Z \sim$20--36),
or fission fragments of the quasi-projectile,
with a possible component of emissions from the neck region 
(light fragments with velocities close to that of the center-of-mass).
Since all products are forward emitted in the center-of-mass reference frame, 
it is reasonable to suppose that quasi-target contributions are negligible.  
As already shown in \cite{Barlini13}, the charge and mass
distributions of the detected products present significant differences
between the $n$-poor and $n$-rich systems, in spite of the fact that the projectile is
the same and the accessible phase space is associated predominantly to
quasi-projectile ejectiles. 
This fact was taken as a proof of isospin diffusion. 
We now want to investigate in how far some differences can be found also in the
staggering of the final yields of fragments.
It is worth noting that being staggering a differential effect between
neighboring nuclei, the detection efficiencies cancel out almost exactly.
Moreover, being the kinematics of the two colliding systems very similar, 
also geometric effects are practically the same in the two sets of data.

\begin{figure}[t]
\begin{center}
\includegraphics[width=9cm] {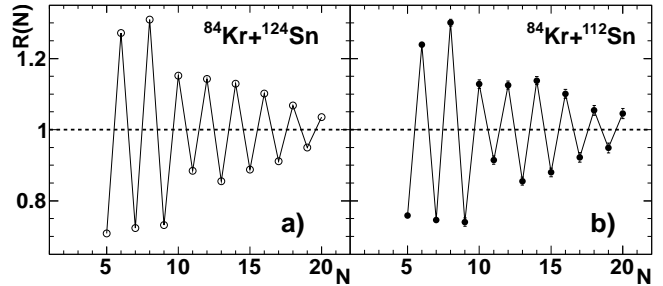} 
\end{center}
 \caption{Staggering as a function of $N$, highlighted by the ratio $R(N)$ 
   for the system (a) $^{84}$Kr+$^{124}$Sn and (b) $^{84}$Kr+$^{112}$Sn,
   both at 35 MeV/nucleon. 
   Bars indicate statistical errors.}\label{fig2}
\end{figure}

To put in quantitative evidence the odd-even staggering one has to remove from
the experimental yields $Y$ the dependence of the smoothed yield 
$\mathcal{Y}$ on varying proton or neutron number of the fragments.
This can be obtained in various ways \cite{Tracy72,Zeitlin97,DAgostino11}.
The treatment of Tracy et al. \cite{Tracy72}, based on a finite difference
method of third order, gives a quantitative measure of the effect and has been
used by most authors.
In this paper we have used a similar procedure, based on the finite differences
of fourth order, that uses five data point and will be described in a
forthcoming paper \cite{metodi_staggering}; 
one advantage is that it avoids using semi-integer values of Z.
We have checked that anyhow the presented results are very little sensitive 
to the particular method used to estimate the smooth behavior of the yield.
For each point of the yield distribution, one can finally build the ratio between
the experimental and the smoothed yields, $R = Y / \mathcal{Y}$, which 
by construction oscillates above and below the line $R=1$ and gives a direct
visual impression of the staggering.

Figures \ref{fig1}(a) and \ref{fig1}(b) display the staggering in $Z$ 
(already visible in the charge distributions of fig. 3 in Ref. \cite{Barlini13})
by means of the ratio $R(Z)$ for the $n$-rich $n$-poor systems, respectively.
The amplitude of the odd-even effect is on average larger for the $n$-poor system, 
thus confirming the findings of previous papers \cite{Yang99,DAgostino11,Lombardo11}. 
Quantitatively the staggering in $Z$ remains of the order of $\approx \pm 10\%$.
For both systems, the staggering is rather pronounced at low-medium $Z$ 
(up to $\sim 20$), then it tends to disappear for higher $Z$ values. 
Around $Z=30$ we observe a renewed increase of the staggering, mainly in the 
$n$-poor system. 
A very similar behavior was observed also in \cite{Casini12}, 
both in inclusive analysis and with some selection of the centrality; 
in that case the studied system was $^{112}$Sn+$^{58}$Ni at 35 MeV/nucleon. 

Thanks to the good isotopic resolution of the FAZIA telescopes, it is here
possible to perform an extensive analysis also for the staggering in $N$, 
for the first time in a rather wide range.  
Figures \ref{fig2}(a) and \ref{fig2}(b) present the staggering in $N$ by means of
the ratio $R(N)$ for the two systems. 
Here the $N$ distribution does not extend beyond $N=20$, because we have 
isotopic resolution up to $Z \approx 20$ (and correspondingly up to 
$N \approx 22$, with the method requiring two points on both sides of each $N$).
This is the limit of our isotopic resolution in the present case. 

The most apparent --and to our knowledge rather new--
feature is that the staggering as a function of $N$ is large
(definitely much larger than that in $Z$),
especially for the lighter fragments where it reaches a rather surprising 
value of $\approx \pm \,30 \%$, and slowly decreases with increasing $N$.
Indeed it strongly
differs from the typical behavior in low-energy fission, where the fission 
fragments usually display a staggering in $N$ weaker than in $Z$
\cite{Siegert76,Schmidt01}.
The second observation is that, at first sight, the behavior of the
staggering in $N$ is very similar in the $n$-rich and $n$-poor systems, and 
this seems in contrast with the conclusions of Lombardo et al. 
\cite{Lombardo11} in lighter systems at 25 MeV/nucleon 
for 4$\leq\!N\!\leq$13.

\begin{figure}[t!]
\begin{center}
\includegraphics[width=9cm] {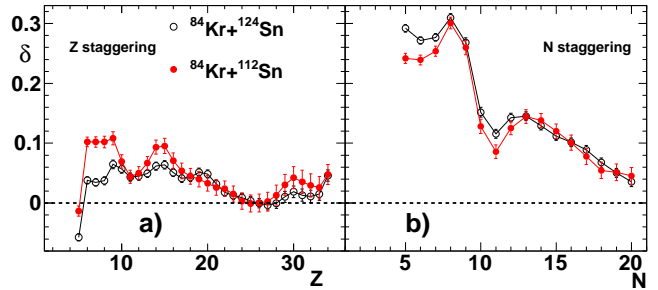} 
\end{center}
\caption{(color online) Parameter $\delta$ as a function (a) of $Z$ and (b) of $N$ for
   final fragments in the  
   collisions $^{84}$Kr+$^{112}$Sn (full symbols),
   and $^{84}$Kr+$^{124}$Sn (open symbols). Bars indicate statistical errors.
}
\label{fig3}
\end{figure}

\begin{table*}[!]
\caption{\label{tab1} Average value of the staggering parameter 
  $\langle \delta \rangle$ as a function of $N$ and $Z$ for the systems
  Kr+Sn of this paper and Ca+Ca \cite{Lombardo11}.
  For Ca+Ca, relative yields and errors are estimated from Fig. 3 
  of Ref. \cite{Lombardo11}.
  For Kr+Sn the averages are evaluated in different ranges of $Z$ and $N$, the
  first one being the same used for the data of Ref. \cite{Lombardo11}.
}
\begin{ruledtabular}
\begin{tabular}{  c   p{15mm}    c       c       c
                p{5mm} p{10mm} p{10mm} p{10mm} 
                p{5mm} p{10mm} p{10mm}
                p{5mm} p{10mm} p{10mm}
}
               System & \multicolumn{1}{c}{Energy} & \multicolumn{3}{c}{$(N/Z)$} 
   & & \multicolumn{3}{c}{~~~~$\langle \delta_Z \rangle$ ($\times 10^3$) }
   & & \multicolumn{2}{c}{~~~~$\langle \delta_N \rangle$ ($\times 10^3$) }
   & & \multicolumn{2}{c}{~~Ratio $\langle\delta_N\rangle/\langle\delta_Z\rangle$}
\\
     &\multicolumn{1}{c}{[MeV/u]} & proj. & targ. & tot. 
       & \multicolumn{2}{r}{$Z=$~6--11~~}
           & \multicolumn{1}{c}{~5--18}
           & \multicolumn{1}{c}{~5--34}
       & \multicolumn{2}{r}{$N=$~6--11~~}
           & \multicolumn{1}{c}{~5--18} 
       & \multicolumn{2}{r}{$Z,N=$~6--11~~~}
           & \multicolumn{1}{c}{~5--18} 
 \\
\\
$^{84}$Kr+$^{112}$Sn &\multicolumn{1}{c}{35} &1.33&1.24&1.28 
   & & \multicolumn{1}{c}{~91$\,\pm\,$4}
     & \multicolumn{1}{c}{~68$\,\pm\,$3}
     & \multicolumn{1}{c}{~53$\,\pm\,$2}
   & & \multicolumn{1}{c}{224$\,\pm\,$4}
     & \multicolumn{1}{c}{173$\,\pm\,$3}
   & & \multicolumn{1}{c}{~2.5$\,\pm\,$0.1}
     & \multicolumn{1}{c}{~2.5$\,\pm\,$0.2}
\\
$^{84}$Kr+$^{124}$Sn &\multicolumn{1}{c}{``} &1.33&1.48&1.42
   & & \multicolumn{1}{c}{~44$\,\pm\,$3}
     & \multicolumn{1}{c}{~40$\,\pm\,$2}
     & \multicolumn{1}{c}{~30$\,\pm\,$1}
   & & \multicolumn{1}{c}{242$\,\pm\,$3}
     & \multicolumn{1}{c}{171$\,\pm\,$2}
   & & \multicolumn{1}{c}{~5.4$\,\pm\,$0.3}
     & \multicolumn{1}{c}{~4.3$\,\pm\,$0.3}
\\
\\
$^{40}$Ca+$^{40}$Ca  &\multicolumn{1}{c}{25} &1.0 &1.0 &1.0
   & & \multicolumn{1}{c}{167~~}
     & \multicolumn{1}{c}{~--} 
     & \multicolumn{1}{c}{~--}
   & & \multicolumn{1}{c}{~71$\,\pm\,$~5}
     & \multicolumn{1}{c}{--~~~~~~} 
   & & \multicolumn{1}{c}{0.43~~}
     & \multicolumn{1}{c}{--}
\\
$^{40}$Ca+$^{48}$Ca &\multicolumn{1}{c}{``} &1.0 &1.4 &1.2 
   & & \multicolumn{1}{c}{~87~~}
     & \multicolumn{1}{c}{~--} 
     & \multicolumn{1}{c}{~--}
   & & \multicolumn{1}{c}{~83$\,\pm\,$10}
     & \multicolumn{1}{c}{~--~~~~~~} 
   & & \multicolumn{1}{c}{0.95~~}
     & \multicolumn{1}{c}{~--}
\\
$^{48}$Ca+$^{48}$Ca &\multicolumn{1}{c}{``} &1.4 &1.4 &1.4 
   & & \multicolumn{1}{c}{~27~~}
     & \multicolumn{1}{c}{~--} 
     & \multicolumn{1}{c}{~--}
   & & \multicolumn{1}{c}{200~}
     & \multicolumn{1}{c}{~--~~~~~~} 
   & & \multicolumn{1}{c}{7.4~~}
     & \multicolumn{1}{c}{~--}
\\
\end{tabular}
\end{ruledtabular}
\end{table*}

As in the method originally proposed by Tracy \cite{Tracy72},
one can use a parameter $\delta(Z)= (-1)^Z (R(Z)-1)$ 
to describe in a quantitative way the behavior of staggering phenomena:
a positive $\delta(Z)$ corresponds to the usual staggering that favors the
production of even $Z$ (or $N$); $\delta(Z)\!\approx\,$0 means absence of any
significant staggering, while negative $\delta(Z)$ indicates a reverse
effect (``anti-staggering'') favoring the production of fragments with odd 
$Z$ (or $N$) values.
The obtained values of the parameter $\delta$ are presented in Fig. \ref{fig3}, 
both for the staggering in $Z$ [part (a)] and in $N$ [part (b)];
full symbols are for the $n$-poor system and open symbols for the $n$-rich one.

The main characteristics, already visible in Figs. \ref{fig1} and
\ref{fig2}, appear even clearer in this presentation:
i) the staggering in $N$ is significantly higher than that in $Z$, by a factor
   of about 3 or more;
ii) the staggering in $N$ is indeed very similar for both systems
(except for the marginal region $N\!\leq$7); 
iii) the staggering in $Z$ tends to disappear above $Z$=20 up to $Z\!\sim\,$28, 
with a sudden clear bump (in spite of the large statistical errors) around
$Z$=30 \cite{Casini12}, which is more pronounced for the $n$-poor system;
iv) the staggering in $Z$ shows some difference between the two systems,
with the $n$-poor system featuring higher $\delta$ below $Z$=10, between $Z$=12
and $Z$=18, and around $Z$=30.
The negative value for $\delta$($Z$=5) in Fig. \ref{fig3} is caused by the
missing $^8$Be, which distorts the needed yield of Be isotopes much more than
the yield of $N\,$=4 isotones.

In fragmentation reactions it was observed \cite{Knott96} that the even-odd 
staggering in $Z$ is reduced for $n$-rich projectiles (like $^{40}$Ar) with
respect to symmetric ones (like $^{36}$Ar). 
Recently Lombardo et al. \cite{Lombardo11} found that also at Fermi energies 
a $n$-rich system has a reduced staggering in $Z$ and an enhanced one in $N$, 
while the opposite happens for a $n$-poor system.
They drew their conclusion on the base of a parameter $S$ (obtained from the 
squared deviations with respect to a polynomial fit to the yield distributions
in the interval 4$\leq\!N\!\leq$13 or 4$\leq\!Z\!\leq$13, see \cite{Lombardo11}) 
that summarizes in a single number the average importance of the staggering in
each system.
Applying that procedure to our case would give too rough an approximation, because
our distributions span a range more than twice as large and hence a simple 
polynomial fit would give a poor description of the smoothed distributions. 
Therefore we prefer to apply our procedure also to their data and present
in Table \ref{tab1} averaged values of the parameter $\delta$, 
obtained in different ranges of $Z$ and $N$.

Our results show that the staggering in $N$ is definitely larger than that in
$Z$, by a factor between 2 and 5.
Concerning the comparison of the two systems, $^{84}$Kr+$^{112}$Sn and 
$^{84}$Kr+$^{112}$Sn, the staggering in $N$ is the same within errors
when evaluated over the full distributions,
thus supporting the visual impression already conveyed by Fig. \ref{fig1} 
(and in seeming contradiction with \cite{Lombardo11}).
However, if only nuclei in the range 4$\leq\!N\!\leq$13 are used for averaging
(as it can be done for the data of \cite{Lombardo11}),
then it appears that also in our case
the $n$-rich system has a slightly enhanced staggering in $N$ 
(0.242$\pm$0.003 vs. 0.224$\pm$0.004),
which is mainly due to the lightest nuclei with $N\!\leq$7.
In contrast, the weaker staggering in $Z$ displays a difference 
of about a factor of 2 between the two systems (in fair agreement with  
\cite{Lombardo11}), that in our case persists almost independently
of the considered range of $Z$.

The last two columns of Table \ref{tab1} give the ratios 
$\langle\delta_N\rangle/\langle\delta_Z\rangle$ between the 
staggering parameters in $N$ and $Z$, evaluated in a common range.
For light fragments ($Z$, $N$ up to 11), the clear prevalence of $N$
staggering over $Z$ staggering is stronger in the $n$-rich system 
than in the $n$-poor one, a fact that can be inferred also from the data 
of \cite{Lombardo11}.
This effect is slightly reduced in the larger range of $Z$, $N$ up to 18.
It is worth noting the systematic dependence of the staggering phenomena on
isospin that is displayed by both experiments,
in spite of the differences in total mass and bombarding energy.
With increasing $N/Z$ of the systems, the decrease of the 
staggering in $Z$ is accompanied by an increase of the staggering in $N$.
As a consequence, the ratio $\langle\delta_N\rangle/\langle\delta_Z\rangle$
evolves from about 0.5 for symmetric matter ($N/Z$=1.0) to about 7
for the very asymmetric case ($N/Z$=1.4).

If one takes a look at the 1-proton (1-neutron) separation energies as a
function of $Z$ ($N$) for various $N$ ($Z$),
one finds a clear staggering, mainly due to pairing effects, but there is
no apparent difference between protons and neutrons.
Tentatively, one may relate the different magnitude 
of the staggering in $Z$ and $N$
to the common assumption that pairing correlations, similarly to 
shell effects, should be washed out with increasing excitation energy.
Proton emission is expected to be more probable in the early steps of the 
evaporation (where the excitation energy is higher) rather than in the last 
ones, unless the system is very $n$-poor as in the case of $^{40}$Ca+$^{40}$Ca.
Therefore proton emission might be less sensitive to pairing effects than 
neutron emission, which is expected to prevail in the last steps, also 
because it is insensitive to the repulsive effect of the Coulomb barrier.
  
\begin{figure}[b!]
\begin{center}
\includegraphics[width=9cm] {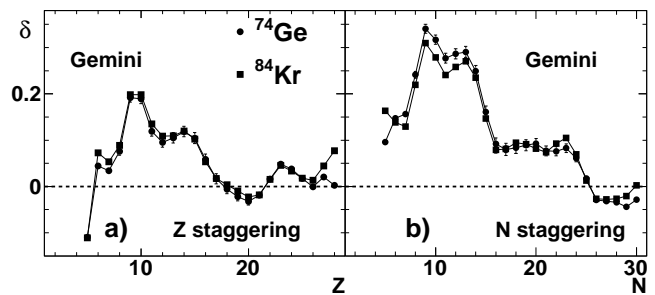} 
\end{center}
\caption{Parameter $\delta$ for the staggering in $Z$ (a) and $N$ (b)
   from {\sc gemini} simulations of the decay $^{84}$Kr (full squares)
   or $^{74}$Ge (full circles) at 2.7 MeV/nucleon of excitation energy.}
\label{gemini1}
\end{figure}

To gain some more insight, we performed calculations with the code {\sc gemini}
for the statistical decay (evaporation and sequential fission followed by 
statistical evaporation) of nuclei with initial excitation energy and spin
corresponding to a semiperipheral collision.
The calculated results are found to be little sensitive
to moderate variations of the input parameters.
The experimental gross features of Fig. \ref{fig3} are 
qualitatively reproduced.
For example, in Figs. \ref{gemini1}(a) and \ref{gemini1}(b) the 
parameter $\delta$ is presented
as a function of $Z$ and $N$ for two decaying nuclei with 2.7 MeV/nucleon of
excitation energy and spin $J=$ 50.
One nucleus (squares) is the $^{84}$Kr projectile, the other (dots) 
is a slightly lighter nucleus of $^{74}$Ge, chosen to simulate 
some pre-evaporative emission, like e.g. in case of mid-velocity 
or pre-equilibrium phenomena.
The magnitude of the $N$ staggering is comparable to that of the experiment
and rapidly decreases with increasing $N$; the magnitude of the $Z$
staggering clearly remains below that of the $N$ staggering. 
A more detailed reproduction of the experimental data is not attempted,
because the initial distribution of decaying primary reaction products is
unknown and cannot be simulated by the decay of a single nucleus with a 
single value of the excitation energy and spin.

\begin{figure}[t!]
\begin{center}
\includegraphics[width=9cm] {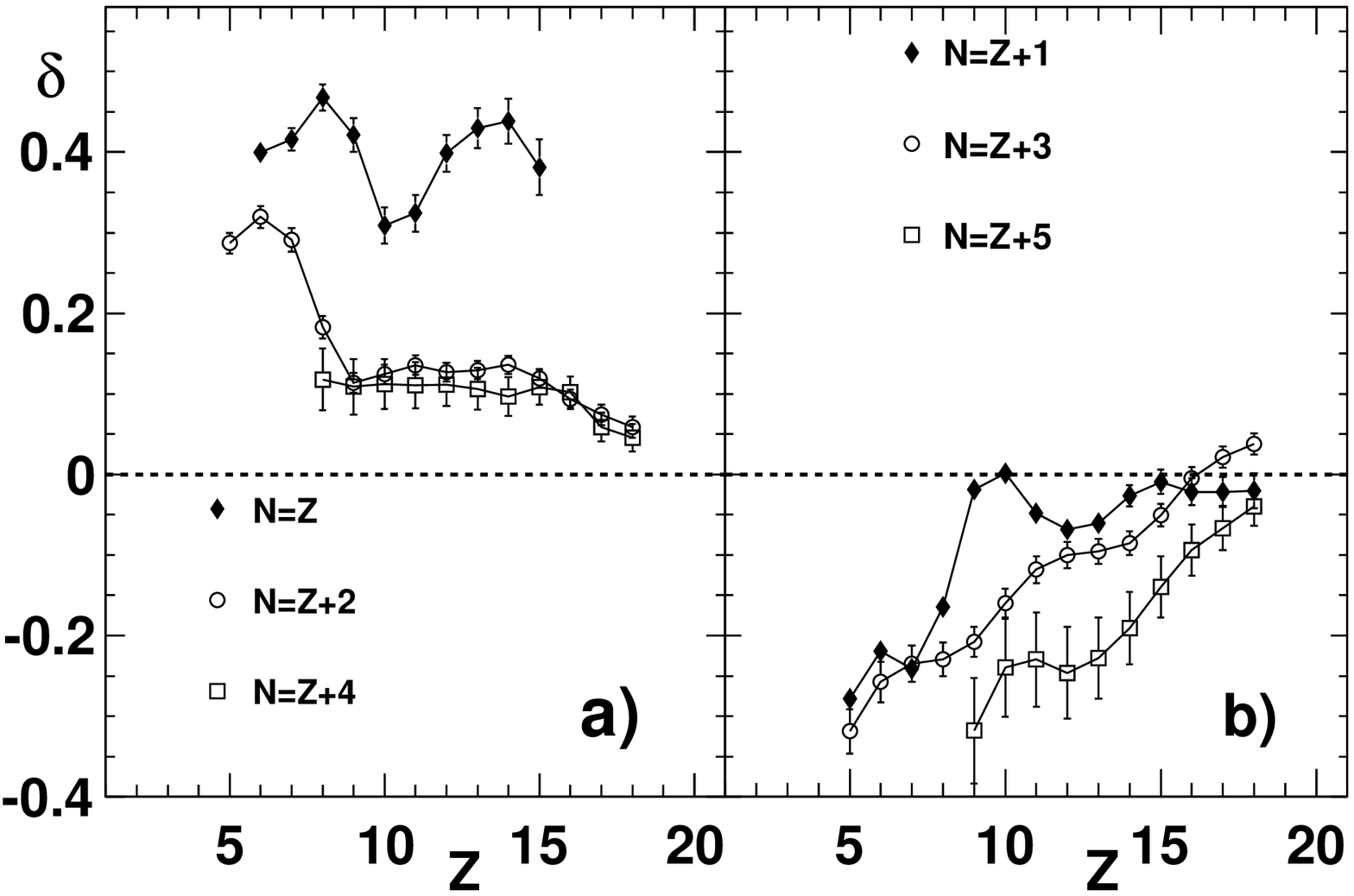} 
\includegraphics[width=9cm] {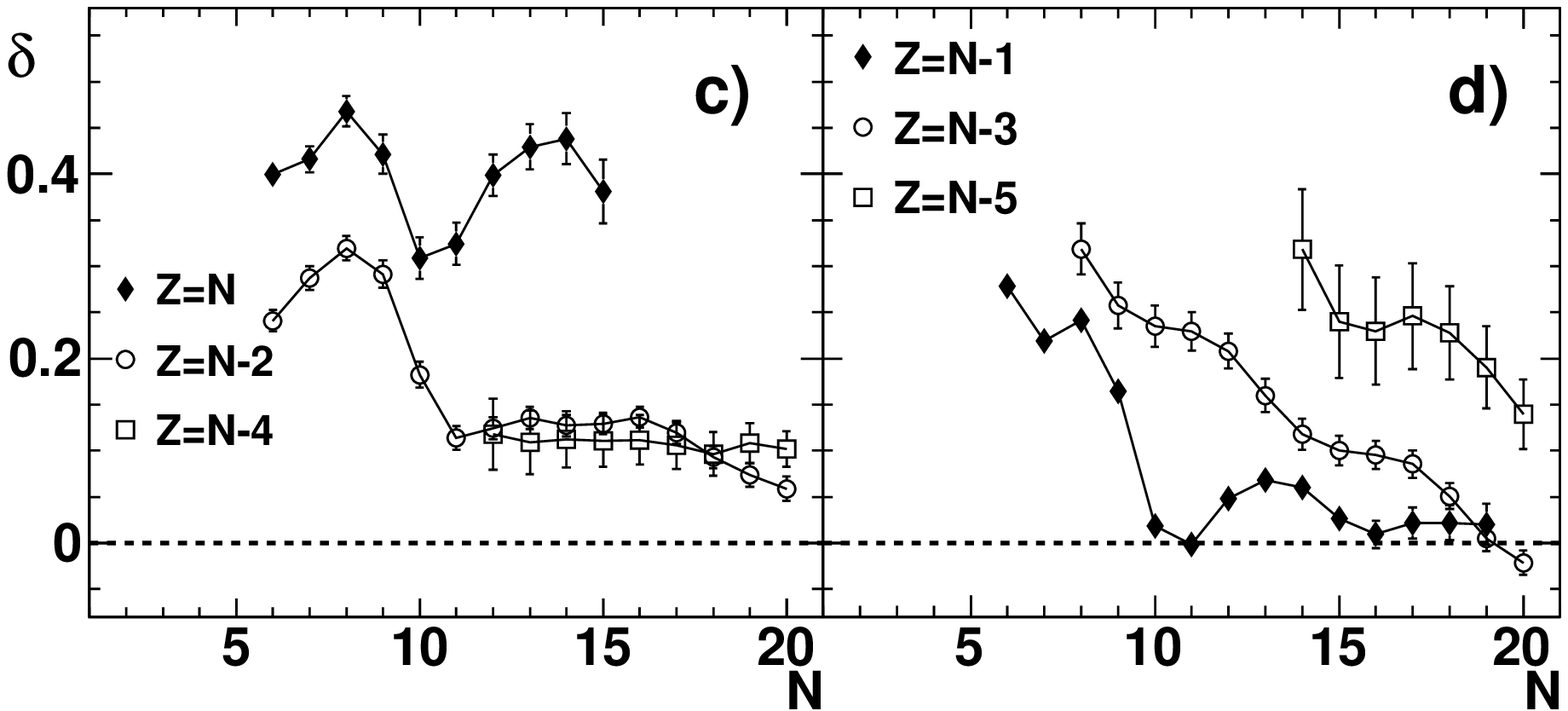} 
\end{center}
\caption{Parameter $\delta$ for the reaction $^{84}$Kr + $^{124}$Sn
  plotted as a a function of $Z$, for even (a) and odd (b) chains of
  neutron excess $N\!-\!Z$
  and as a function of $N$, again for even (c) and
  odd (d) chains of $N\!-\!Z$.
  Note the larger vertical scale with respect to Fig. \ref{fig3}(a).
}
\label{fig5}
\end{figure}

In the literature, the staggering in $Z$ has been often looked at 
for chains of constant neutron excess $N\!-\!Z$
\cite{Ricciardi04,DAgostino11,winkenbauer_arXiv1303}.
Figure \ref{fig5} shows this presentation of the data for the system 
$^{84}$Kr+$^{124}$Sn.
Similar results are obtained for the other system $^{84}$Kr+$^{112}$Sn.
In the upper left panel there are the chains with even $N\!-\!Z$ and 
in the upper right one the chains with odd $N\!-\!Z$.
One sees that the $N\!=\!Z$ chain displays by far the largest positive
staggering, namely a strong enhancement of even $Z$ with respect to the 
neighboring odd values resulting in positive values of $\delta$.
The effect for the other chains with even $N\!-\!Z$ is definitely smaller.
In contrast, chains with odd $N\!-\!Z$ seem to display a negative staggering
(or ``antistaggering''), namely a depression of the yields of even $Z$
(negative values of $\delta$),
which appears to be stronger for nuclei with larger values of $N\!-\!Z$.

A similar qualitative behavior (although with much larger uncertainties) 
is observed in Fig. 4 of \cite{Ricciardi04} for the fragmentation of 
1 GeV/nucleon $^{238}$U in a titanium target and an 
even quantitative agreement is found with the data of
Fig. 11 of \cite{Napolitani07}, concerning the spallation of 1 GeV/nucleon 
$^{136}$Xe in a liquid hydrogen target.

The general behavior observed in Fig.\ref{fig5} can be understood 
simply from the fact that 
there is staggering {\it both} in $N$ and $Z$ 
(i.e., even $N$ and $Z$ values are enhanced and odd ones are depressed)
and the effect is larger in $N$ than in $Z$.
The staggering is thus intensified for the even $N\!-\!Z$ chains of
Fig. \ref{fig5}(a), which are formed only by even-even nuclei (benefiting from
both enhancements) and odd-odd nuclei (depressed by both effects).
In case of odd $N\!-\!Z$ chains, the nuclei are always odd-even or even-odd 
and therefore the staggering in $N$ and $Z$ works in opposite directions.
The net result is that even $Z$ are depressed due to the prevalent 
effect of odd $N$ contributions and, conversely, odd $Z$ are 
enhanced due to the prevalent effect of even $N$: the net result is the 
moderate {\it seeming} ``anti-staggering'' visible in Fig. \ref{fig5}(b).

The same data can be plotted as a function of the neutron content $N$ of 
the fragments, as shown in the lower panels of Fig. \ref{fig5}.
The points are exactly the same as in the upper panels, there are just
horizontal shifts for the various chains and an additional change of sign 
for all chains corresponding to odd-A nuclei in Fig. \ref{fig5}(d) 
with respect to Fig. \ref{fig5}(b).
Therefore the {\it seeming\, } antistaggering in $Z$, 
commonly observed for odd mass nuclei, is an artifact of the selection: 
in reality the production of final fragments is intensified for even $Z$ and 
even $N$ nuclei, with a more pronounced effect for the $N$ ``pairing''.
This is at variance with what was usually observed in low-energy fission.

\section{SUMMARY AND CONCLUSIONS}

In summary, we have investigated the odd-even staggering effects in the 
yields of fragments produced in two reactions with the same beam of $^{84}$Kr 
at 35 MeV/nucleon and two different targets, one $n$-rich ($^{124}$Sn) and 
one $n$-poor ($^{112}$Sn).
The data were collected by the FAZIA Collaboration by means of a 
telescope located close to the projectile grazing angle.
The high resolution of the telescope allowed us to obtain good isotopic 
identification for all ions in the wide range up to $Z\approx20$.

The staggering was studied for complex fragments emitted in the phase space of
the quasi-projectile (residues, fission products, midvelocity products). 
For the present analysis, the usual parameter $\delta$ \cite{Tracy72}, which
allows to perform quantitative comparisons among different sets of data,
has been slightly modified \cite{metodi_staggering}.
The staggering of medium-light fragments has been extensively analyzed as a
function of both the atomic number $Z$ and the neutron number $N$,
for the first time over a rather wide range.
It is found that, for a given reaction, the staggering in $N$ is definitely
larger than that in $Z$. 
In agreement with other authors \cite{Yang99,DAgostino11,Lombardo11}, 
we observe in the $n$-poor system a larger staggering in $Z$ with respect 
to the $n$-rich one, while the staggering in $N$ is in general rather similar, 
being slightly larger only for the lightest fragments
produced in the $n$-rich system.
However the difference between the two systems is smaller for the staggering 
in $N$ and varies with the considered range in $N$.
Simulations with the {\sc gemini} code \cite{CharityPhysRevC.82.014610}
qualitatively reproduce the larger effect 
for $N$ staggering.

The staggering in $Z$ for selected values of the neutron excess $N\!-\!Z$ 
presents features similar to those already reported in literature
\cite{Ricciardi04,winkenbauer_arXiv1303,DAgostino11}. 
Qualitatively they arise from the interplay between staggering in $Z$ and $N$.
The production of final fragments is intensified for even 
values of both $Z$ and $N$, with the latter dominating over the former.
The reason why the staggering in $N$ is larger than that in $Z$ 
and their dependence on isospin
remain for the moment obscure and deserve further investigations.
They will strongly benefit from 
the future availability of unstable radioactive beams and from the development
of high-resolution detectors, covering large solid angles and coupled 
with setups capable of a good characterization of the events.

\begin{acknowledgments}
Many thanks are due to the crew of the Superconducting Cyclotron,
in particular D. Rifuggiato, for providing a very good
quality beam, and to the staff of LNS for continuous support.
The authors wish to warmly thank also R. J. Charity for discussions about 
the {\sc Gemini} code.
The support of the detector and mechanical workshops of the Physics Department 
of Florence is gratefully acknowledged.
Funding was received from the European Union Seventh Framework Programme 
FP7(2007-2013) under Grant Agreement No. 262010-ENSAR.
We acknowledge support by the Foundation for Polish Science MPD program,
co-financed by the European Union within the European Regional Development
Fund.
\end{acknowledgments}

\bibliography{referenze}

\end{document}